\documentclass[a4paper, oneside, twocolumn, notitlepage, 10pt]{extarticle_ecoc}
\usepackage{ecoc}

\usepackage[acronym]{glossaries}
\usepackage{bm}
\usepackage{siunitx}
\usepackage{upgreek}
\usepackage{nicefrac}
\usepackage{kitcolors}
\usepackage{pgfplots}
\usepackage{siunitx}

\usepackage[font=footnotesize]{subcaption}

\usepackage{todonotes}

\usetikzlibrary{fit, positioning, calc, angles, arrows.meta}
\pgfplotsset{compat=newest}

\pgfplotsset{
  discard if/.style 2 args={
    x filter/.code={
      \edef\tempa{\thisrow{#1}}
      \edef\tempb{#2}
      \ifx\tempa\tempb
      
      \fi
    }
  },
  discard if not/.style 2 args={
    x filter/.code={
      \edef\tempa{\thisrow{#1}}
      \edef\tempb{#2}
      \ifx\tempa\tempb
      \else
      
      \fi
    }
  }
}

\let\Re\relax
\let\Im\relax
\DeclareMathOperator{\Re}{Re}
\DeclareMathOperator{\Im}{Im}

\let\j\relax
\DeclareMathOperator{\j}{j}
\definecolor{cb-1}{HTML}{4477AA}
\definecolor{cb-2}{HTML}{EE6677}
\definecolor{cb-3}{HTML}{228833}
\definecolor{cb-4}{HTML}{CCBB44}
\definecolor{cb-5}{HTML}{66CCEE}
\definecolor{cb-6}{HTML}{AA3377}
\definecolor{cb-7}{HTML}{BBBBBB}

\pgfplotscreateplotcyclelist{cb list}{
    cb-1,cb-2,cb-3,cb-4,cb-5,cb-6,cb-7
}

\glsdisablehyper

\newacronym{bps}{BPS}{blind phase search}
\newacronym{rpn}{RPN}{residual phase noise}
\newacronym{awgn}{AWGN}{additive white Gaussian noise}
\newacronym{gs}{GS}{geometrical shaping}
\newacronym{qam}{QAM}{quadrature amplitude modulation}
\newacronym{snr}{SNR}{signal to noise ratio}
\newacronym{bce}{BCE}{binary cross entropy}
\newacronym{bmi}{BMI}{bitwise mutual information}
\newacronym{gcs}{GCS}{geometric constellation shaping}
\newacronym{pgcs}{pGCS}{parameterizable geometric constellation shaping}
\newacronym{gmi}{GMI}{generalized mutual information}
\newacronym{mi}{MI}{mutual information}
\newacronym{e2e}{E2E}{end-to-end}
\newacronym{cpe}{CPE}{carrier phase estimation}
\newacronym{llr}{LLR}{log-likelikood ratio}
\newacronym{nn}{NN}{neural network}
\newacronym{tx}{Tx}{transmitter}
\newacronym{rx}{Rx}{receiver}
\newacronym{fec}{FEC}{forward error correction}
\newacronym{bmd}{BMD}{bit-metric decoder}
\newacronym{ff-nn}{FF-NN}{feed-forward neural network}
\newacronym{ber}{BER}{bit error rate}
\newacronym{dsp}{DSP}{digital signal processing}
\newacronym{vv}{V\&V}{Viterbi \& Viterbi}

\newcommand\extrafootertext[1]{%
    \bgroup
    \renewcommand\thefootnote{\fnsymbol{footnote}}%
    \renewcommand\thempfootnote{\fnsymbol{mpfootnote}}%
    \footnotetext[0]{#1}%
    \egroup
}

\colorlet{KITColor1}{kit-blue100}
\colorlet{KITColor2}{kit-orange100}
\colorlet{KITColor3}{kit-maigreen100}

\pgfplotsset{
  discard if/.style 2 args={
    x filter/.code={
      \edef\tempa{\thisrow{#1}}
      \edef\tempb{#2}
      \ifx\tempa\tempb
      
      \fi
    }
  },
  discard if not/.style 2 args={
    x filter/.code={
      \edef\tempa{\thisrow{#1}}
      \edef\tempb{#2}
      \ifx\tempa\tempb
      \else
      
      \fi
    }
  }
}

\makeatletter
\DeclareRobustCommand{\rvdots}{%
  \vbox{
    \baselineskip4\p@\lineskiplimit\z@
    \kern-\p@
    \hbox{.}\hbox{.}\hbox{.}
  }}
\makeatother

\newsavebox\neuralnetwork
\sbox{\neuralnetwork}{%
		\begin{tikzpicture}[
      >=stealth,
      scale=.5,
      every node/.append style={transform shape},
      remember picture,
      ]
      \tikzset{Source1b/.style={rectangle, draw=black, thick, minimum width=0.05cm, minimum height=1.2cm, rounded corners=0.5mm}}
      \tikzset{Source3/.style={rectangle, draw, thick, minimum width=0.6cm, minimum height=0.45cm, rounded corners=0.5mm}}

      \tikzset{OnehotNode/.style={circle, thick, draw,minimum width=0.1cm}}
      \tikzset{ReLUNode/.style={circle,thick,draw,fill=black!10!white}}
      \tikzset{MZMNode/.style={circle,thick,draw,fill=black!30!white}}
   \node (serial) at (3.6,0) {};

    \def\k{0}
    \def\ki{1}
        \node [OnehotNode] (S\ki1) at ($(0,0.5)+(0,\k)$) {};
        \node [OnehotNode] (S\ki2) at ($(0,-0.5)+(0,\k)$) {};
		\node at ($(0,0)+(0,\k)$) {$\rvdots$};
		\draw [->] ($(S\ki1)+(-0.4cm,0)$) -- (S\ki1);
		\draw [->] ($(S\ki2)+(-0.4cm,0)$) -- (S\ki2);

		\node [ReLUNode] (H\ki11) at ($(1,1.5)+(0,\k)$) {};
		\node [ReLUNode] (H\ki12) at ($(1,0.5)+(0,\k)$) {};
		\node [ReLUNode] (H\ki13) at ($(1,-0.5)+(0,\k)$) {};
		\node [ReLUNode] (H\ki14) at ($(1,-1.5)+(0,\k)$) {};
		\node at ($(1,0)+(0,\k)$) [anchor=center] {$\rvdots$};

		\foreach\j in {1,2,3,4}{
		  \draw [->] (S\ki1) -- (H\ki1\j); \draw [->] (S\ki2) -- (H\ki1\j);
		};

		\node [ReLUNode] (H\ki21) at ($(2,1.5)+(0,\k)$) {};
		\node [ReLUNode] (H\ki22) at ($(2,0.5)+(0,\k)$) {};
		\node [ReLUNode] (H\ki23) at ($(2,-0.5)+(0,\k)$) {};
		\node [ReLUNode] (H\ki24) at ($(2,-1.5)+(0,\k)$) {};
		\node at ($(2,0)+(0,\k)$) [anchor=center] {$\rvdots$};

		\foreach\j in {1,2,3,4}{
		  \foreach\i in {1,2,3,4}{
		    \draw [->] (H\ki1\j) -- (H\ki2\i);
		  };
		};

		\node [MZMNode] (M\ki1) at ($(3,1)+(0,\k)$) {};
		\node [MZMNode] (M\ki2) at ($(3,0)+(0,\k)$) {};
		\node [MZMNode] (M\ki3) at ($(3,-1)+(0,\k)$) {};
		\draw [<-] ($(serial.west)+(0,\k)+(0,1)$) -- (M\ki1.east);
		\draw [<-] ($(serial.west)+(0,\k)$) -- (M\ki2.east);
		\draw [<-] ($(serial.west)+(0,\k)+(0,-1)$) -- (M\ki3.east);

		\node at ($(3,-0.5)+(0,\k)$) [anchor=center] {$\rvdots$};
		\node at ($(3,+0.5)+(0,\k)$) [anchor=center] {$\rvdots$};
		\foreach\j in {1,2,3,4}{
		  \foreach\i in {1,2,3}{
		    \draw [->] (H\ki2\j) -- (M\ki\i);
		  };
		};
\end{tikzpicture}%
}

\begin{document}
\selectlanguage{american}    


\title{Optimized Geometric Constellation Shaping for Wiener Phase Noise Channels with Viterbi-Viterbi Carrier Phase Estimation}%


\author{
    Andrej Rode, Wintana Araya Gebrehiwot, Shrinivas Chimmalgi, and Laurent
    Schmalen%
    \vspace*{-1ex}
}

\maketitle                  


\begin{strip}
 \begin{author_descr}

   Communications Engineering Lab (CEL), Karlsruhe Institute of Technology (KIT), 
   \textcolor{blue}{\uline{rode@kit.edu}}%
   \vspace*{-1ex}
 \end{author_descr}
\end{strip}

\setstretch{1.1}
\renewcommand\footnotemark{}
\renewcommand\footnoterule{}
\interfootnotelinepenalty=10000 


\begin{strip}
  \begin{ecoc_abstract}%
  The \gls{vv} algorithm is well understood for QPSK and 16-QAM, but modifications are required for higher-order modulation formats. We present an approach to extend the standard \gls{vv} algorithm for higher-order modulation formats by modifying the transmit constellation with geometric constellation shaping.  
    \textcopyright2023 The Author(s)%
    \vspace*{-1.5ex}
  \end{ecoc_abstract}
\end{strip}


\section{Introduction}
\extrafootertext{This work has received funding from the European Research
  Council (ERC) under the European Union's Horizon 2020 research and innovation
  programme (grant agreement No. 101001899).}%
\extrafootertext{This paper is a preprint of a paper submitted to ECOC 2023 and is subject to Institution of Engineering and Technology Copyright. If accepted, the copy of record will be available at IET Digital Library.}%
In recent years, the optimization of geometric constellation shaping for communication systems impaired by Wiener phase noise has been a topic of great interest~\cite{dzieciolGeometricShaping2D2021,jovanovicGradientfreeTrainingAutoencoders2021,jovanovicEndtoendLearningConstellation2022,rodeEndtoendOptimizationConstellation2023}. The optimization is typically performed using a state-of-the-art auto-encoder approach in a symbol-wise or bit-wise manner~\cite{cammererTrainableCommunicationSystems2020}. To incorporate effects of the communication channel and subsequent carrier phase estimation into the optimization, either a residual phase noise process~\cite{jovanovicGradientfreeTrainingAutoencoders2021,jovanovicEndtoendLearningConstellation2022,dzieciolGeometricShaping2D2021} or a Wiener phase noise process with a carrier phase estimation~\cite{rodeEndtoendOptimizationConstellation2023} has been included in the training. In the first case, the influence of the constellation on the operation of the carrier phase estimation is neglected and only the impairment of the signal at the output of the carrier phase estimation is considered in the training. Compared to the second approach, this simplifies the optimization and training. In the second case, to perform \gls{e2e} optimization with gradient descent the operations of the carrier phase estimation have to be implemented in a differentiable manner. In~\cite{rodeEndtoendOptimizationConstellation2023}, we have introduced a differentiable \gls{bps} to optimize constellations geometrically. In this work, we explore this approach for the \gls{vv} algorithm and introduce geometrically optimized 64-ary constellations for a Wiener phase noise channel with \gls{vv} \gls{cpe}. We modify the \gls{vv} algorithm to include a learnable and differentiable partitioning in the geometrical constellation shaping optimization process. 

\section{Feed-forward Carrier Phase Estimation}
For high-rate coherent optical communication receivers, the choice of the \gls{cpe} algorithm is driven by the symbol rate compared to the processing speed of \gls{dsp}. As the symbol rate is multiple orders of magnitude higher than the signal processing rate of \gls{dsp}, the use of feedback \gls{cpe}---common in communication receivers e.g. for wireless communications---is not possible. A popular feed-forward \gls{cpe} implementation is the \gls{bps}~\cite{pfauHardwareefficientCoherentDigital2009}, which is widely used in modern optical communication receivers. Another feed-forward \gls{cpe} implementation is the \gls{vv} \gls{cpe}~\cite{viterbiNonlinearEstimationPSKmodulated1983a}, where computational complexity is traded off for performance. 

To perform \gls{cpe} with the \gls{bps},  the squared distance to each constellation point needs to be calculated for each received complex symbol. For square \gls{qam}, the distance all of the constellation points must be calculated for 1/4 of the test phases. For higher-order constellations, this increases the computational complexity significantly. For \gls{vv}-based \gls{cpe}, the complexity does not scale with the number of constellation points, but with $\mu$, since the $\mu$-th power of each received complex symbol is computed followed by averaging and an estimation of the phase. For square \gls{qam} constellations, usually $\mu=4$ is chosen due to the four-fold symmetry. Since we apply geometric shaping, the parameter $\mu$ can be chosen freely to increase or decrease the symmetry in the constellation with respect to square \gls{qam}. This is of particular interest for the robustness to cycle slips, as a lowered $\mu$ reduces the rotational symmetry of the constellation and the region of phase unambiguity increases. Additionally, a smaller $\mu$ leads to a less complex \gls{cpe}.
\begin{figure*}[tbh]
\tikzset{MUL/.style={draw,circle,append after command={
      [shorten >=\pgflinewidth, shorten <=\pgflinewidth,]
      (\tikzlastnode.north west) edge (\tikzlastnode.south east)
      (\tikzlastnode.north east) edge (\tikzlastnode.south west)
    }
  }
}
\tikzset{ADD/.style={draw,circle,append after command={
      [shorten >=\pgflinewidth, shorten <=\pgflinewidth,]
      (\tikzlastnode.north) edge (\tikzlastnode.south)
      (\tikzlastnode.east) edge (\tikzlastnode.west)
    }
  }
}
\tikzstyle{box}=[rectangle,draw=black, minimum size=7mm, inner sep=2mm, font=\footnotesize, align=center]
\tikzstyle{node}=[circle, draw=black, minimum size=5mm]
\tikzstyle{connection}=[-{Latex}]
\tikzset{%
  cblock/.style    = {draw, rectangle, minimum height = 3, minimum width = 6,rounded corners=1mm},
  lblock/.style = {draw, rectangle,minimum height=10, minimum width=2,
    rounded corners=1mm},
  operation/.style = {draw, minimum height= 1.5em, circle},
}
\centering
\begin{tikzpicture}[font=\small, thick, x=3mm, y=3mm]
        \node [lblock, minimum width=3] (txnet) at (0,0) {\begin{tabular}{c} Tx-NN\\\usebox{\neuralnetwork}\end{tabular}};
        \node [left=2 of txnet,align=left] (source) {$\begin{pmatrix}b_{k,1}\\ b_{k,2}\\ \vdots\\ b_{k,m}\end{pmatrix}$};
        \node [operation, ADD, right=2.5 of txnet] (awgn_add) {};
        \node [below=1.2 of awgn_add, text depth=0.8em, text height=1em] (awgn_param) {$n_k$};

        \node [operation, MUL, right=1.5 of awgn_add] (phase_mult) {};
        \node [below=1.2 of phase_mult, text depth=0.8em, text height=1em] (pn_param) {$\mathrm{e}^{\mathrm{j}{\phi}_k}$};
		\node [above=1.2 of phase_mult] (pn_param_def) {$\Delta\phi_k \sim \mathcal{N}(0,\sigma_\upphi^2$)};
        \node [lblock,right=2 of phase_mult] (vv){V\&V CPE};
		\node [lblock,right=2 of vv] (complexrx) {\begin{tabular}{c}{Rx-NN}\\\usebox{\neuralnetwork}\end{tabular}};
		\node [right=1.2 of complexrx,align=left] (sink) {$\begin{pmatrix}\hat{L}_{k,1}\\ \hat{L}_{k,2}\\ \vdots \\ \hat{L}_{k,m}\end{pmatrix}$};
		\draw [connection] (source) -- (txnet);
		\draw [connection] (awgn_param) -- (awgn_add);
		\draw [connection] (pn_param) -- (phase_mult);
        \draw [connection] (txnet) -- node[above]{$x_k$} (awgn_add);
        \draw [connection] (awgn_add) -- (phase_mult);
        \draw [connection] (phase_mult) -- node[above]{$z_k$} (vv);
        \draw [connection] (vv) -- (complexrx);
        \draw [connection] (complexrx) -- (sink);

\end{tikzpicture}
    \caption{System model to learn geometrically shaped constellations for \acrshort{vv}-based \acrshort{cpe}.}
    \label{fig:system_model_vv}
\end{figure*}
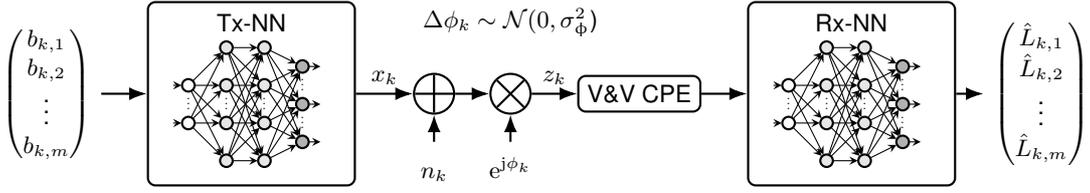
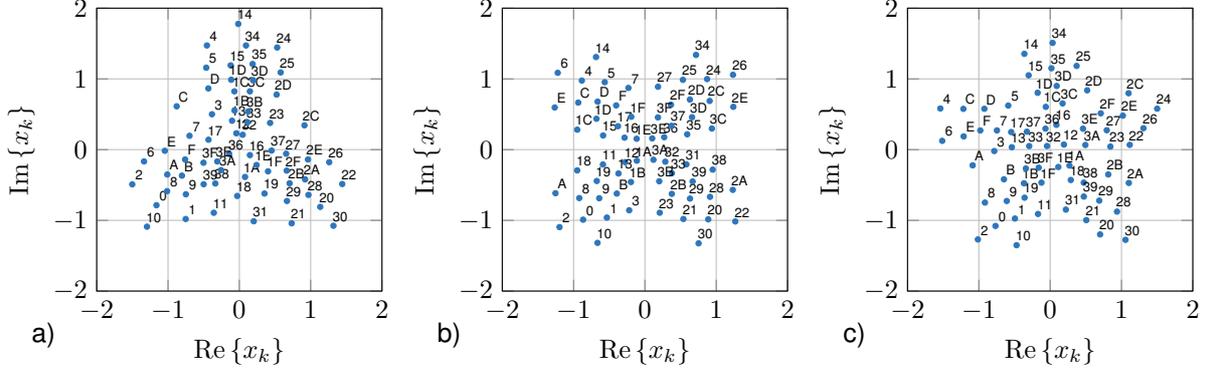
\begin{figure*}[tbh]
    \centering
    \begin{subfigure}{\textwidth/3}
    \begin{tikzpicture}
    \tikzstyle{expr}=[font=\footnotesize]
    \tikzstyle{block}=[expr,rectangle, draw, thick, minimum size=30pt,
minimum height=15pt, minimum width=30pt, inner sep=3pt, rounded corners=1, fill=white]%
    \begin{axis}[
      xlabel={$\Re\left\{x_k\right\}$},
      ylabel={$\Im\left\{x_k\right\}$},
      height=\textwidth,
      width=\textwidth,
      xmin=-2,
      xmax=2,
      ymin=-2,
      ymax=2,
      grid=both,
      ]
      \addplot[
      only marks,
      mark=*,
      mark size=1pt,
      color=KITColor1,
      coordinate style/.from={black,scale=0.5,xshift=5pt},
      nodes near coords,
      point meta=explicit symbolic,
      ]
      table[col sep=tab, meta=label]
      {data/64QAM/constellation_sym3.txt};
    \end{axis}
    \node[below = 3mm,xshift=-0.7cm] {a)};
  \end{tikzpicture}
  \end{subfigure}%
      \begin{subfigure}{\textwidth/3}
    \begin{tikzpicture}
    \tikzstyle{expr}=[font=\footnotesize]
    \tikzstyle{block}=[expr,rectangle, draw, thick, minimum size=30pt,
minimum height=15pt, minimum width=30pt, inner sep=3pt, rounded corners=1, fill=white]%
    \begin{axis}[
      xlabel={$\Re\left\{x_k\right\}$},
      ylabel={$\Im\left\{x_k\right\}$},
      height=\textwidth,
      width=\textwidth,
      xmin=-2,
      xmax=2,
      ymin=-2,
      ymax=2,
      grid=both,
      ]
      \addplot[
      only marks,
      mark=*,
      mark size=1pt,
      color=KITColor1,
      coordinate style/.from={black,scale=0.5,xshift=5pt},
      nodes near coords,
      point meta=explicit symbolic,
      ]
      table[col sep=tab, meta=label]
      {data/64QAM/constellation_sym4.txt};
    \end{axis}
    \node[below = 3mm,xshift=-0.7cm] {b)};
  \end{tikzpicture}
  \end{subfigure}%
        \begin{subfigure}{\textwidth/3}
    \begin{tikzpicture}
    \tikzstyle{expr}=[font=\footnotesize]
    \tikzstyle{block}=[expr,rectangle, draw, thick, minimum size=30pt,
minimum height=15pt, minimum width=30pt, inner sep=3pt, rounded corners=1, fill=white]%
    \begin{axis}[
      xlabel={$\Re\left\{x_k\right\}$},
      ylabel={$\Im\left\{x_k\right\}$},
      height=\textwidth,
      width=\textwidth,
      xmin=-2,
      xmax=2,
      ymin=-2,
      ymax=2,
      grid=both,
      ]
      \addplot[
      only marks,
      mark=*,
      mark size=1pt,
      color=KITColor1,
      coordinate style/.from={black,scale=0.5,xshift=5pt},
      nodes near coords,
      point meta=explicit symbolic,
      ]
      table[col sep=tab, meta=label]
      {data/64QAM/constellation_sym5.txt};
    \end{axis}
    \node[below = 3mm,xshift=-0.7cm] {c)};
  \end{tikzpicture}
  \end{subfigure}%
    \caption{Learned constellations for the \gls{vv}-based \gls{cpe} for $m=\SI{6}{bit/symbol}$ and a) $\mu=3$, b) $\mu=4$ and c) $\mu=5$. Bit labels are obtained by converting the bit vectors to their hexadecimal representation, e.g., $(0, 1, 1, 1, 1, 1) \equiv 1F$}
    \label{fig:constellations_without_partitioning}
\end{figure*}

\Gls{cpe} based on the \gls{vv} algorithm are not well-suited for higher order \gls{qam}, as not all constellation points are located on the symmetry lines, and therefore the inclusion of those points will result in reduced phase estimation performance. Previous approaches to improve \gls{vv}-based \gls{cpe} use a system of partitioning the received constellation symbols based on their amplitude into multiple classes~\cite{fatadinLaserLinewidthTolerance2010b,bilalMultistageCarrierPhase2014}. In the second step, the \gls{vv} algorithm is only applied to symbols in a certain class. For the case of 16-\gls{qam}, only the outer- and innermost points are selected for the phase estimation.
A solution that includes more points to improve the phase estimate of the \gls{vv} \gls{cpe} rotates the received symbols to the symmetry lines~\cite{bilalCarrierPhaseEstimation2015a}. Another approach to improve the \gls{cpe} performance uses a multi-stage approach, where the first stage performs a coarse phase estimate at a lower complexity, while the following stages can refine the phase estimate using pre-compensated complex symbols~\cite{farukDigitalSignalProcessing2017}. Adding additional stages to the phase estimation increases latency and computational complexity. Therefore, we look at approaches that leave a single-stage \gls{vv} algorithm in place and instead adapt the transmit constellation to improve the performance.

\section{Geometric Constellation Shaping with Modified \gls{vv} Algorithm}
To improve the performance of the \gls{vv} algorithm for higher-order constellations, we apply geometric constellation shaping using the bitwise auto-encoder approach. The \gls{vv} algorithm to obtain the phase estimate at time $k$ for the received complex symbol $z_k$
\begin{equation}
    \varphi_{k,\mathrm{est}} = \frac{1}{\mu}\mathrm{unwrap}\left(\arg\left(\sum_{k'=k-K}^{k+K}z_{k'}^\mu\right)\right),
\end{equation}
 is already a differentiable operation.
 Therefore we can insert the \gls{vv} in the \gls{e2e} training with the bitwise auto-encoder as shown in \autoref{fig:system_model_vv}.
 To perform geometric shaping, while still following the approach of partitioning, the previously presented partitioning approach~\cite{fatadinLaserLinewidthTolerance2010b} needs to be modified. Hard partitioning selecting certain symbols based on the amplitude will leave the optimization without any useful gradients and the partitioning cannot be directly included in the optimization. We introduce a novel, differentiable  selection function that carries out the partition, with the goal to include many constellation points in the constellation optimization and phase estimate. The proposed selection function is
\begin{equation}
    z'_{k,l} = \sigma\left(s_l(|z_k| - \theta_{0,l})\right) \sigma\left(s_l(\theta_{1,l} - |z_k|)\right) \frac{z_k}{|z_k|},
\end{equation}
which will apply a selection on each received complex symbol $z_k$ to obtain $z'_{k,l}$. The function is parameterized by the trainable parameters $s_l$, $\theta_{0,l}$, and $\theta_{1,l}$ and we apply  the non-linear Softplus activation function $\sigma$. To have multiple partitioning rings, we define multiple functions indexed by $l \in \{0,\ldots,L-1\}$ and calculate the phase estimate according to 
\begin{align}
    \Tilde{z}_{k,\mathrm{avg}} &= \sum_{k'=k-K}^{k+K}\left(\sum_{l=0}^{L-1} z'_{k',l}\right)^{\mu}\label{eq:new_phase_estimate} \\
    \varphi_{k,\mathrm{est},\mathrm{mod}} &= \frac{1}{\mu}\mathrm{unwrap}\left(\arg\left(\Tilde{z}_{k, \mathrm{avg}}\right)\right).
\end{align}
For $L=0$ no partitioning is applied on the received symbols.
In~\eqref{eq:new_phase_estimate}, a weighted average is calculated over $2K+1$ neighboring phasors.  The phase estimate $\varphi_{k,\mathrm{est},\mathrm{mod}}$ for the symbol at time step $k$ is then obtained by taking the complex argument and performing phase unwrapping to have a continuous phase variation.
In a subsequent step, we perform averaging across the neighboring phase estimates included in the partitioning to obtain phase estimates for constellation symbols that were not included in the partitioning step.

\section{Results}

\begin{figure}[t!]
    \centering
    \begin{tikzpicture}
\pgfplotsset{
sym3/.style ={KITColor1,dashed,mark=*},
sym4/.style ={KITColor2,solid,mark=square*},
sym5/.style ={KITColor3,dashdotted,mark=triangle*}
}
  \begin{axis}[
    xlabel={Linewidth (\si{kHz})},
    ylabel={BMI (\si{bit/symbol})},
    ymin=3.5,
    ymax=5.5,
    xmin=0,
    xmax=1000,
    grid=both,
    width=0.8\columnwidth,
    height=\columnwidth,
    mark size=1pt,
    legend style={at={(axis cs:1050,3.5)},anchor=south west,nodes={transform shape},font=\footnotesize,text width={3.3em},legend cell align={left}},
    label style={font=\small},
        cycle multi list={
            cb list\nextlist
            {thick, solid, mark=*},{thick, dotted,mark options={solid}, mark=x},{thick, dashed, mark options={solid}, mark=o},
            {thick, densely dashed, mark options={solid}, mark=triangle*},
            {thick, densely dotted, mark options={solid}, mark=pentagon*}\nextlist
            }
    ]
    \addlegendimage{thick, solid, mark=*, black} \addlegendentry[minimum height=2.3em]{$\mu=4$ $L=1$}
    \addlegendimage{thick, dotted, mark options={solid}, mark=x, black} \addlegendentry[minimum height=2.3em]{$\mu=4$ $L=0$}
    \addlegendimage{thick, dashed, mark options={solid}, mark=o, black} \addlegendentry[minimum height=2.3em]{$\mu=5$ $L=0$}
    \addlegendimage{thick, densely dashed, mark options={solid}, mark=triangle*, black} \addlegendentry[minimum height=2.3em]{$\mu=3$ $L=0$}
    \addlegendimage{thick, densely dotted, mark options={solid}, mark=pentagon*, black} \addlegendentry[minimum height=2.3em]{QAM $L=2$};
    \foreach\s in{15.00,17.00,19.00}{
    \addplot+[
    ]
    table[discard if not={snr}{\s}, x expr={\thisrow{linewidth}/1000}, y=mean, col sep=space, y error=stddev]
    {data/64QAM/nn_vv_sym4_p1.txt};
    \addplot+[
    ]
    table[discard if not={snr}{\s}, x expr={\thisrow{linewidth}/1000}, y=mean, col sep=space,y error=stddev]
    {data/64QAM/nn_vv_g4755ymp_sym4.txt};
    \addplot+[
    ]
    table[discard if not={snr}{\s}, x expr={\thisrow{linewidth}/1000}, y=mean, col sep=space,y error=stddev]
    {data/64QAM/nn_vv_9j5m2l72_sym5.txt};
    \addplot+[
    ]
    table[discard if not={snr}{\s}, x expr={\thisrow{linewidth}/1000}, y=mean, col sep=space, y error=stddev]
    {data/64QAM/nn_vv_mxq0bg44_sym3.txt};

    \addplot+[
    ]
    table[discard if not={snr}{\s}, x expr={\thisrow{linewidth}/1000}, y=mean, col sep=space,y error=stddev]
    {data/64QAM/qam_part2.txt};}
  \end{axis}
  \begin{axis}[
    hide axis,
    ymin=3.5,
    ymax=6,
    xmin=0,
    xmax=1000,
    grid=both,
    width=0.8\columnwidth,
    height=\columnwidth,
    mark size=1pt,
    reverse legend,
    legend style={at={(axis cs:1050,6)},anchor=north west,nodes={transform shape},font=\footnotesize, text width={3.3em}},
    label style={font=\small},
    legend cell align={left},
        cycle multi list={
            cb list
            }
  ]

        \foreach\s/\stext in{15.00/15\,dB,17.00/17\,dB,19.00/19\,dB}{
        \addplot+[no marks, very thick] coordinates {(100,4)};
        \addlegendentryexpanded{\stext}
     }
     \addlegendimage{empty legend}\addlegendentry{SNR}
  \end{axis}
\end{tikzpicture}
    \caption{Validation results for transmission with \acrshort{vv}-based \acrshort{cpe} without partitioning and a genie-aided cycle slip compensation for $m=\SI{6}{bit/symbol}$.}
    \label{fig:performance_without_partitioning}
\end{figure}
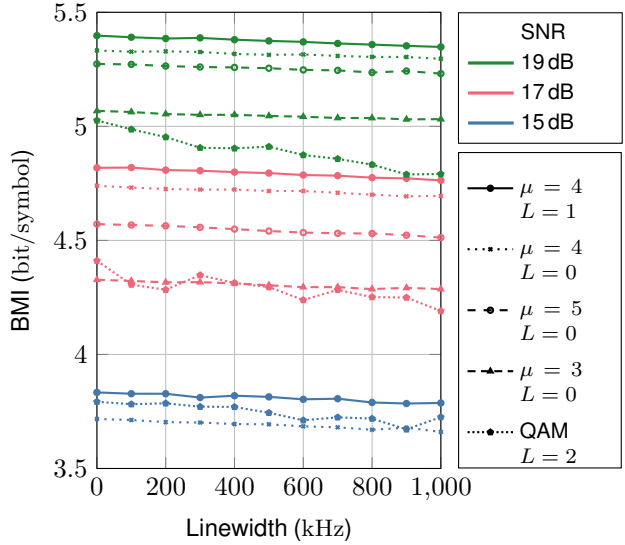

\begin{figure}[tb]
    \centering
    \begin{tikzpicture}
    \pgfplotsset{colormap={kit-cm}{color={KITblue} color={KITorange}}}
    \tikzstyle{expr}=[font=\footnotesize]
    \tikzstyle{block}=[expr,rectangle, draw, thick, minimum size=30pt,
minimum height=15pt, minimum width=30pt, inner sep=3pt, rounded corners=1, fill=white]%
    \begin{axis}[
      xlabel={$\Re\left\{x_k\right\}$},
      ylabel={$\Im\left\{x_k\right\}$},
      height=0.8\columnwidth,
      width=0.8\columnwidth,
      xmin=-2,
      xmax=2,
      ymin=-2,
      ymax=2,
      grid=none,
      view={0}{90},
      unbounded coords=jump,
      point meta rel=per plot,
      colormap name=kit-cm,
      colorbar,
      colorbar style={
        font=\small,
        ytick={0.5,0.75,1},
        ylabel={partition},
      },
      ]
      \addplot3
      [surf,
       shader=flat,
       mesh/cols=50,
       point meta min=0.5,
       point meta max=1,
       ]
      table [col sep=tab, x=real, y=imag, z=partition]{data/64QAM/constellation_sym4_p1_partition.txt};
      \addplot[
      scatter,
      scatter/use mapped color={
        draw=mapped color,
        fill=mapped color,
      },
      mark=*,
      mark size=1pt,
      scatter/use mapped color={
        draw=mapped color,
        fill=mapped color,
      },
      draw=none,
      coordinate style/.from={white,scale=0.5,xshift=5pt},
      nodes near coords,
      point meta=explicit symbolic,
      ]
      table[col sep=tab, meta=label]
      {data/64QAM/constellation_sym4_p1.txt};
    \end{axis}
  \end{tikzpicture}
    \caption{Learned constellations for the \gls{vv}-based \gls{cpe} with learned partitioning for $m=\SI{6}{bit/symbol}$ and $\mu=4$. Bit labels are obtained by converting the bit vectors to their hexadecimal representation, e.g., $(0, 1, 1, 1, 1, 1) \equiv 1F$}
    \label{fig:constellations_with_partitioning}
\end{figure}
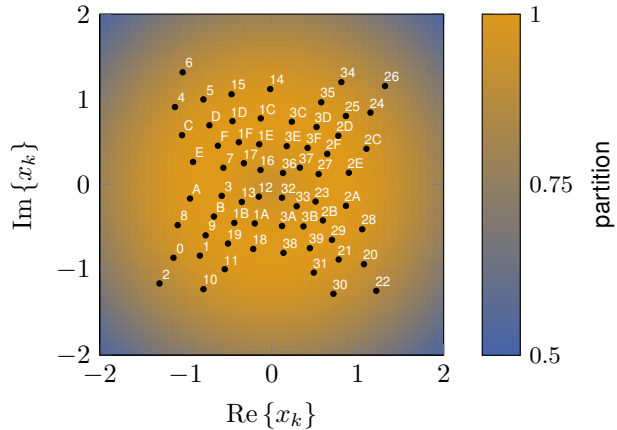

Performing \gls{gcs} without partitioning for $\mu \in {2,3,4}$, we obtain constellations with a significant amount of constellation points located close to the corresponding symmetry lines. We display the constellations for \gls{vv} \gls{cpe} without partitioning in \autoref{fig:constellations_without_partitioning}. With a bit-wise autoencoder approach, the bit labeling is also included in the optimization and we print the obtained bit labels for each constellation symbol in \autoref{fig:constellations_without_partitioning}.

The training and validation were performed at a symbol rate of $R_{\mathrm{S}} = \SI{32}{GBaud}$. For training, the \gls{snr} was fixed to \SI{20}{dB}, and the laser linewidth was fixed to \SI{100}{kHz}. For the validation, we use a genie-aided cycle slip compensation to simulate a system operating with a phase offset which can be corrected unambiguously with a \gls{vv} \gls{cpe}. This can be achieved with a first stage \gls{cpe}  employing a pilot-based~\cite{bilalMultistageCarrierPhase2014} or pilot-less~\cite{rodeEndtoendOptimizationConstellation2023} approach. Comparing the performance of the geometrically optimized constellations for varying $\mu$ in \autoref{fig:performance_without_partitioning}, we can observe that for lower \glspl{snr} the system with $\mu=4$ shows the best performance and for high \glspl{snr} also the constellation and system with $\mu=5$ 
approaches the performance of $\mu=4$. From this, we conclude that the classical choice of $\mu=4$ for square \gls{qam} also provides the best performance for a geometrically optimized constellation. The constellation trained with $L=1$ for partitioning shown in \autoref{fig:constellations_with_partitioning} shows a consistently better performance compared to the constellation without partitioning. Since the partitioning is performed in a way, where most of the symbols are included in the \gls{vv} \gls{cpe}, the increased performance might stem from the additional averaging step which is performed after the partitioning.

\section{Conclusions}
We present an approach to optimize the \gls{gcs} of higher order modulations for \gls{vv}-based \gls{cpe}. We firstly introduce our \gls{e2e} optimization system based on the bitwise auto-encoder and then present a modified \gls{vv} \gls{cpe}, which applies a differentiable partitioning on the constellation to further improve the performance. We compare the results of our approaches in terms of the \gls{bmi} and find that optimization of the \gls{gcs} allows for the application of the \gls{vv} \gls{cpe} on higher order modulation formats without any partitioning. Application of partitioning further improves the performance, while adding some additional computational complexity to the \gls{dsp}. Furthermore, an analysis of the exponent $\mu$ used in the \gls{vv} algorithm shows, that for 64-ary constellations and optimized \gls{gcs} a value of $\mu=4$ provides the best performance.

\clearpage
\newpage
\printbibliography

\vspace{-4mm}

\end{document}